\documentclass[12pt]{article}

\usepackage{cite}

\usepackage{amssymb,latexsym}

\usepackage{graphics,epsfig}

\usepackage{color}

\let\a=\alpha \let\b=\beta \let\g=\gamma \let\d=\delta \let\e=\epsilon
\let\z=\zeta  \let\th=\theta  \let\k=\kappa
\let\l=\lambda \let\m=\mu \let\n=\nu \let\x=\xi \let\p=\pi 
\let\s=\sigma   \let\f=\phi  
 
       \let\D=\Delta \let\Th=\Theta 
\let\X=\Xi  \let\S=\Sigma  \let\Y=\Psi
 
\let\la=\label  
  
\def\nn{\nonumber} \def\bd{\begin{document}} \def\ed{\end{document}}
\def\ds{\documentstyle} \let\fr=\frac \let\bl=\bigl \let\br=\bigr
\let\Br=\Bigr \let\Bl=\Bigl
\let\bm=\bibitem
\let\na=\nabla
\def\tU{{\widetilde U}}
\let\pa=\partial \let\ov=\overline
\def\ie{{\it i.e.\ }}
\newcommand{\be}{\begin{equation}}
\newcommand{\ee}{\end{equation}}
\def\ba{\begin{array}}
\def\ea{\end{array}}
\def\ft#1#2{{\textstyle{{\scriptstyle #1}\over {\scriptstyle #2}}}}
\def\fft#1#2{{#1 \over #2}}
\def\F#1#2{{ F_{#1}^{(#2)} }}
\def\cF#1#2{{ {\cal F}_{#1}^{(#2)} }}

\def\R{{\bf R}}
\def\sst#1{{\scriptscriptstyle #1}}
\def\oneone{\rlap 1\mkern4mu{\rm l}}
\def\e7{E_{7(+7)}}
\def\td{\tilde}
\def\wtd{\widetilde}
\def\im{{\rm i}}
\def\bog{Bogomol'nyi\ }
\newcommand{\ho}[1]{$\, ^{#1}$}
\newcommand{\hoch}[1]{$\, ^{#1}$}
\newcommand{\bea}{\begin{eqnarray}}
\newcommand{\eea}{\end{eqnarray}}
\newcommand{\ra}{\rightarrow}
\newcommand{\lra}{\longrightarrow}
\newcommand{\Lra}{\Leftrightarrow}
\newcommand{\ap}{\alpha^\prime}
\newcommand{\bp}{\tilde \beta^\prime}
\newcommand{\cB}{{\cal B}}
\newcommand{\cO}{{\cal O}}
\newcommand{\vecx}{\vec{x}}
\newcommand{\vecy}{\vec{y}}
\newcommand{\vecp}{\vec{p}}
\newcommand{\vecq}{\vec{q}}
\newcommand{\tr}{{\rm tr} }
\newcommand{\Tr}{{\rm Tr} }
\newcommand{\NP}{Nucl. Phys. }

\newcommand{\cL}{{\cal L}}
\newcommand{\cA}{{\cal A}}
\newcommand{\cT}{{\cal T}}
\newcommand{\cD}{{\cal D}}
\newcommand{\cH}{{\cal H}}

\def\sst#1{{\scriptscriptstyle #1}}
\def\0{{\sst{(0)}}}
\def\1{{\sst{(1)}}}
\def\2{{\sst{(2)}}}
\def\3{{\sst{(3)}}}
\def\4{{\sst{(4)}}}
\def\5{{\sst{(5)}}}
\def\6{{\sst{(6)}}}
\def\7{{\sst{(7)}}}
\def\8{{\sst{(8)}}}
\def\9{{\sst{(9)}}}
\def\p{{\sst{(p)}}}
\def\q{{\sst{(q)}}}
\def\ve{\varepsilon}
\def\vf{\varphi}
\def\F{\Phi}
\def\wg{\wedge}

\def\thb{\bar{\theta}}
\def\Thb{\bar{\Theta}}
\def\barp{\bar{p}}
\def\barq{\bar{q}}
\def\barc{\bar{c}}
\def\bard{\bar{d}}
\def\e{\epsilon}

\def \bi{\bibitem}
\def \la {\label}

\def \l {\lambda}
\def\foot{\footnote}
\def \tl  {{\tilde \l}}
\def \sql {{\sqrt \l}}
\def \adss {$AdS_5 \times S^5$\ }
\newcommand{\rf}[1]{(\ref{#1})}
\def \ov {\over}

\def\th{\theta}
\def\Th{\Theta}
\def\vth{\vartheta}
\def\btheta{{\bar\theta}}
\def\ttheta{{{\tilde\theta}}}
\def\bttheta{{{\bar\ttheta}}}
\def\vth{\vartheta}

\def\ra{\rightarrow}
\def\N{\nabla}
\def\F{{\cal F}}
\def\uM{\underline{M}}
\def\uA{\underline{A}}
\def\uN{\underline{N}}
\def\uP{\underline{P}}
\def\ua{\underline{a}}
\def\ub{\underline{b}}
\def\uc{\underline{c}}
\def\ud{\underline{d}}
\def\ue{\underline{e}}
\def\uf{\underline{f}}
\def\ui{\underline{i}}
\def\uj{\underline{j}}
\def\uk{\underline{k}}
\def\ul{\underline{l}}
\def\ual{\underline{\alpha}}
\def\ube{\underline{\beta}}
\def\um{\underline{m}}
\def\un{\underline{n}}
\def\up{\underline{p}}
\def\uq{\underline{q}}
\def\ur{\underline{r}}
\def\us{\underline{s}}
\def\umu{\underline{\mu}}
\def\unu{\underline{\nu}}
\def\ula{\underline{\l}}
\def\uka{\underline{\k}}
\def\usi{\underline{\s}}
\def\urh{\underline{\r}}
\def\cc{\circ}
\def\eqv{\equiv}

\def\ni{\noindent}

\def\Ep{E^{{}^{(+)}}}
\def\Em{E^{{}^{(-)}}}

\def\Mp{M^{{}^{(+)}}}
\def\Mm{M^{{}^{(-)}}}

\def \ha{{1\ov 2}}

\def\r{\rho}

\def\Y{{\rm Y}}
\def\X{{\rm X}}
\def\tY{\tilde{\rm Y}}
\def\tX{\tilde{\rm X}}
\def\dY{\dot{\rm Y}}
\def\dX{\dot{\rm X}}

\def \J {\mathcal{J}}
\def \del {\partial}

\def\dF{\dot{F}}
\def\dG{\dot{G}}
\def\df{\dot{f}}
\def \E {{\cal E}}
\def \S {{\cal S}}
\def \J {{\cal J}}

\def\ms{\mathcal{S}}
\def\mj{\mathcal{J}}
\def\soj{\fr{\ms}{\mj}}
\def \R {{\bf R}}
\def \om {\omega}
\def \bE {\bar E}
\def \x {{\cal X}}

\def \bi{\bibitem}
\def \la {\label}

\def \l {\lambda}
\def\foot{\footnote}
\def \tl  {{\tilde \l}}
\def \sql {{\sqrt \l}}
\def \adss {$AdS_5 \times S^5$\ }
\def \ov {\over}

\def \varpi {{\rm w}}

\def\thb{\bar{\theta}}
\def\Thb{\bar{\Theta}}
\def\mb{\bar{\m}}
\def\ab{\bar{\a}}
\def\zb{\bar{z}}
\def\psib{\bar{\psi}}
\def\barp{\bar{p}}
\def\barq{\bar{q}}
\def\barc{\bar{c}}
\def\bard{\bar{d}}
\def\e{\epsilon}
\def\wb{\bar{w}}
\def\lb{\bar{\l}}
\def\Jb{\bar{J}}
\def\Nb{\bar{N}}
\def\Zb{\bar{Z}}
\def\pab{\bar{\pa}}

\def\At{\tilde{A}}
\def\Bt{\tilde{B}}
\def\Ct{\tilde{C}}
\def\Dt{\tilde{D}}
\def\Et{\tilde{E}}
\def\Ft{\tilde{F}}
\def\Gt{\tilde{G}}
\def\Ht{\tilde{H}}
\def\Mt{\tilde{M}}
\def\Rt{\tilde{R}}
\def\at{\tilde{a}}
\def\bt{\tilde{b}}
\def\ct{\tilde{c}}
\def\dt{\tilde{d}}
\def\et{\tilde{e}}
\def\ft{\tilde{f}}
\def\htil{\tilde{h}}
\def\gt{\tilde{g}}
\def\mt{\tilde{\mu}}
\def\nt{\tilde{\nu}}
\def\pht{\tilde{\f}}

\def\rht{\tilde{\rho}}

\def\asth{\hat{*}}
\def\phh{\hat{\phi}}

\def\bA{{\bf A}}

\def\ola{\overleftarrow}
\def\ora{\overrightarrow}
\def\alt{\tilde{\a}}

\def\eh{\hat{e}}
\def\eph{\hat{\e}}
\def\ph{\hat{p}}
\def\alh{\hat{\a}}
\def\beh{\hat{\b}}
\def\gah{\hat{\g}}
\def\Fh{\hat{F}}
\def\muh{\hat{\m}}
\def\nuh{\hat{\n}}
\def\thh{\hat{\th}}
\def\rhh{\hat{\r}}
\def\dh{\hat{d}}
\def\ih{\hat{i}}
\def\jh{\hat{j}}
\def\kh{\hat{k}}
\def\deh{\hat{\d}}
\def\wh{\hat{w}}
\def\lah{\hat{\l}}
\def\Ah{\hat{A}}
\def\Ch{\hat{C}}
\def\Omh{\hat{\Omega}}

\def\xh{\hat{x}}

\def\ps{\rlap{\, /}\;\,p }
\def\ks{\rlap{\, /}\;\,k }

\def\gym{g_{YM}}

\def\adot{\dot{a}}
\def\bdot{\dot{b}}
\def\bpa{\bar{\pa}}

\def\pr{\prime}
\def\ssk{\medskip}

\begin{document}

\overfullrule=0pt
\parskip=2pt
\parindent=12pt
\headheight=0in \headsep=0in \topmargin=0in
\oddsidemargin=0in

\vspace{ -3cm}
\thispagestyle{empty}

 \vspace{0.1cm}

\setcounter{equation}{0}
\setcounter{footnote}{0}
\setcounter{section}{0}

\begin{center}

{\Large\bf  ADM reduction of Einstein action and black hole entropy}

\vskip 0.8cm

 \vspace{.5cm}


\vspace{0.5cm}
I. Y. Park
\\



\vspace{0.1cm}
{\it Department of Applied Mathematics,
Philander Smith College 
                               \\
Little Rock, AR 72223, USA \\
inyongpark05@gmail.com
}

\end{center}

 \vspace{0.1cm}

 \begin{abstract}
We reduce the 4D Einstein-Hilbert action to a constant-radius hypersurface of foliation. The resulting theory is a scalar theory defined on a 3D hypersurface of the original black hole background, and has an exponential potential. Once the the hypersurface is located at the Schwarzschild radius, the 3D theory
is effectively reduced to a 2D Liouville type theory. We compute {the entropy associated with the hypersurface intrinsic degrees of freedom}, and show that its leading order reproduces the Bekenstein-Hawking area law. 
The subleading terms come in logarithm/inverse powers of the area.

\end{abstract}
\newpage

\section{Introduction}

Based on the earlier works \cite{Sato:2002kv,Sato:2003ky,Sato:2004ic} (see \cite{Parry:1993mw,Darian:1997mp,Fukuma:2000bz,Shiromizu:2003dr} for related discussions), a new Kaluza-Klein scheme in which the bulk spacetime was viewed as a foliation of hypersurfaces along one
of the coordinates, ``$r$" was proposed in \cite{Hatefi:2012bp}. The scheme - called the ADM reduction\footnote{As discussed in \cite{Hatefi:2012bp} and more recently in \cite{Park:2013vpa,Park:2013bma,Park:2014mba} (which have appeared several months after the first version of the present work), ADM reduction should provide a promising route to derivation of the AdS/CFT-type dualities from the first principle.} - focuses on the  dynamics of the selected hypersurface, which should capture aspects of the bulk physics. The selected coordinate should play a special role, and this naturally suggests a description of dynamics where $r$ plays the role of ``time" coordinate. 

Although the ADM reduction procedure is a variation of the standard Kaluza-Klein reduction, it is significantly different from the latter. The standard reduction reduces a theory ``on" a manifold, and one gets a lower dimensional low energy theory in the ``transverse" dimensions. In contrast, the ADM reduction reduces the original theory directly to a hypersurface of interest. (In the present case, the 4D theory is reduced to a 3D theory defined in $(t,\th,\vf)$ space.) 
Also, virtual boundary effects are important in general as analyzed, e.g., in \cite{Park:2013vpa}. We will comment on them further below.

Applied to a 5D supergravity theory, the ADM reduction led to the worldvolume (the selected leaf) theory of the D3 brane \cite{Sato:2002kv}\cite{Hatefi:2012bp}.
Here we apply the scheme to 4D pure Einstein gravity, and reduce it to a fixed point in the radial coordinate $r$. We show that the reduction leads to a 3D scalar theory description of the hypersurface selected.
Interestingly, the scalar theory takes the form of a 3D analogue of Liouville theory  with an exponential potential.

In the big picture, the ADM reduction scheme proposed in \cite{Hatefi:2012bp} is to describe aspects of the original theory in terms of the lower dimensional theory that results from dimensionally reducing the original theory to a hypersurface of foliation.\footnote{In the original bulk point of view, the reduced theory can be viewed to describe the dynamics of a set of modes with certain symmetry if the reduced direction is an isometry.} The first step in ADM reduction is ADM decomposition.
The ADM reduction that we consider in this work is a relatively simple procedure that does not require the 
full machinery of ADM decomposition; below we will consider a reduction ansatz in the form of 
a block-diagonal decomposition,
\bea
 ds_4^2   \equiv g_{\bar{\mu}\bar{\nu}}dx^{\bar{\mu}}dx^{\bar{\nu}} \equiv n^2(r,x) dr^2+h_{\m\n}(r,x)dx^\m dx^\n
\la{ADMd2q}
\eea
Compared with the standard ADM decomposition (see, e.g., \cite{tool}
for a review), the shift vector is absent. Therefore, \rf{ADMd2q} should be viewed as a combination of the ADM decomposition and
a gauge-fixing in which the shift vector is set to zero.
Only the dynamics associated with fluctuations along the radial direction will be
considered. (The rationale behind this reduction will be discussed below.) Then fluctuations will become dependent on the time and angle coordinates $(t,\th,\vf)$, and eventually, the system is rendered only of a scalar field in the fixed metric background $h_{\m\n}$.
 The scalar field - which we denote by $\f(t,\th,\vf)$ below - is related to
the radial component of the metric in a particular way that is presented in the next section.

In other words, the scalar theory describes a branch of the moduli space of the Einstein-Hilbert action, and
should capture certain aspects of the 4D gravity theory. One may say that the particular branch of the moduli space of the original theory admits a ``dual" description, which is the scalar theory in the present case.
It seems, therefore, that the idea of ``duality branch by branch \cite{Hatefi:2012sy}" is at work here too. In particular, it is plausible for the entropy of the BH system \cite{Bekenstein:1973ur}\cite{Hawking:1976de} 
 to be captured by the scalar theory. Discussions on microscopic origin of the black hole entropy can be 
 found, e.g., in \cite{'tHooft:1984re,Bombelli:1986rw,Srednicki:1993im}. 

One subtle issue is whether one should employ the finite temperature field theory technique
to describe the dynamics of the hypersurface since the black hole has Hawking temperature. We believe that one can employ the zero-temperature field theory setup, and further comment on this issue in the conclusion.

\vspace{.2in}
The rest of the paper is organized as follows.
In section 2, we implement the ADM reduction of 4D Einstein-Hilbert action.
{The goal of this section is to obtain the action of the field $\f(\th,\vf)$, \rf{liouvilletype0}. (Although it is not a direction pursued in the present work, one may consider ADM reduction along $\th$ or $\vf$ direction. If both directions are reduced, one would get an action of a field $\Phi(t,r)$.\footnote{This task has now been carried out in \cite{Park:2013vpa}.})
Although the actual reduction is carried out in the Lagrangian formulation, it is the 
Hamiltonian formulation that provides an intuitive and physical picture. 
After briefly reviewing the Hamiltonian formulation of ``$r$-evolution", we demonstrate that the Schwarzschild metric can be obtained as a solution of this Hamiltonian formulation. 
When the hypersurface is taken to be the event horizon, only the spatial directions survive; the dimensions of the
scalar theory are effectively reduced to two dimensions, and therefore, the theory becomes a Liouville type theory. (Liouville theory was obtained in \cite{Solodukhin:1998tc} in the context of the standard Kaluza-Klein reduction of the 4D Einstein-Hilbert action. Related works include \cite{Cvitan:2002cs,Giacomini:2004tt}.) Compared with the standard form of Liouville theory, the theory obtained has several relatively minor differences. The computation of the entropy of the resulting 2D theory is undertaken in section 3 in which
the area scaling of the entropy is established - in support of the proposal in - up to the issues of 
renormalization and non-perturbative corrections. 
The entropy obtained displays the same pattern that was observed in \cite{Kaul:2000kf} (Entropy corrections were considered also in \cite{Zaslavsky:1997kp}\cite{Mann:1997hm}, earlier and \cite{Sen:2012dw} later\footnote{We thank S.H. Yi for pointing out this reference.}).
We conclude with summary and future directions. The Appendix A contains a discussion of entropy of a quantum field theory system at zero temperature. Details of the reduction can be found in Appendix B.

\section{ADM reduction of 4D Einstein action\la{ADM}}

In this section, we carry out ADM reduction of the 4D Einstein gravity. The ADM reduction 
that we consider in this work is a fairly straightforward procedure that does not require the 
full machinery of ADM decomposition but only its block-diagonal version. Roughly speaking, we leave 
arbitrary degrees of freedom to the $(r,r)$-component of the 4D metric $g_{\umu \unu}$, writing it as $g_{rr}=g_{0rr}+\d g_{rr}$, and view $\d g_{rr}$ -{which is essentially the $\f$ field}- as the field that governs the dynamics of the hyersurface.

The actual reduction is carried out in the Lagrangian formulation for technical simplicity. However, it is the 
Hamiltonian formulation that provides an intuitive picture;
we start by briefly reviewing the Hamiltonian formulation of ``$r$-evolution" in section \ref{hamil}.
We check in section \ref{metric rescaling} that the Schwarzschild metric can be obtained as a solution of the Hamiltonian formulation of $r$-evolution. In section \ref{metric rescaling},
we carry out ADM reduction of the 4D Einstein-Hilbert action (this reduction might be somewhat similar in  spirit to those of \cite{Biswas:2002nk}\cite{Artsukevich:2008vy}) and obtain a Liouville-type theory.

\subsection{Review of Hamiltonian formulation \la{hamil}}

We start with the Einstein-Hilbert action
\bea
S=\fr1{G}\int d^4 x\sqrt{-g}\, R^\4  \la{EH}^{•}
\eea
The 4D coordinate $x^{\umu}$ consists of the radial coordinate $r$ and the 3D coordinate $x^\m$:
\bea
x^{\umu}=r,x^\m\quad \m=0,1,2
\eea
Let us consider the following block-diagonal reduction ansatz for the solution,
\bea
 ds_4^2 
       &=& n^2 dr^2+h_{\m\n}dx^\m dx^\n
\la{ADMd2}
\eea
It is more proper to include the shift vector; with it, the steps below will be slightly modified but with the same conclusion. (However, it is crucial to omit the shift vector in section \ref{metric rescaling} where the reduction is carried out: it turns out that once the shift vector is included, not only the $\f$-fluctuations but also the fluctuations of $h_{\m\n}$ must be included for consistency of the reduction.)
The 4D Einstein-Hilbert action now takes
\bea
\int d^4x\, \sqrt{-g^{(4)}}R^{(4)}=\int dr\int d^3x\, n\sqrt{-h} 
\left(\fr1{4n^2}(h^{\m\n}\pa_r h_{\m\n})^2-\fr1{4n^2}(\pa_r h_{\m\n})^2+R^{(3)}\right)
  \la{action}
\eea
With the lagrangian given by
\bea
\cL=n\sqrt{-h} \left(\fr1{4n^2}(h^{\m\n}\pa_r h_{\m\n})^2-\fr1{4n^2}(\pa_r h_{\m\n})^2+R^{(3)}\right)  \la{L}
\eea
the corresponding Hamiltonian is
\bea
\cH &=& p^{\m\n}\pa_r h_{\m\n}-\cL  \nn\\
  &=& -n\sqrt{-h}\,R^\3+\fr{n}{\sqrt{-h}}\Big(- p_{\m\n}^2+\fr12  (p_\m^\m)^2\Big)
\label{hamiltonian}
\eea
where $p_{\m\n}$ is the canonical momentum, and $ \nabla_\m $ is
the 3D covariant derivative associated with the hypersurface under consideration.
The field equations/constraints that follow from \rf{hamiltonian} are
\bea
-\dot{p}^{\m \n} &=& n\sqrt{-h}\,G^{\m \n}+\fr12 \fr{n}{\sqrt{-h}} (p^{\r\s}p_{\r\s}-\fr12 p^2)h^{\m \n}
-2\fr{n}{\sqrt{-h}} (p_\r^\m p^{\n \r}-\fr12 p p^{\m \n})  \nn\\
    &&-\sqrt{-h}(\N^{(\n} \N^{\m)} n-h^{\m \n}\N^\r \N_\r n)   \label{mtmdr}
\eea
\bea
- R^\3+\fr1{(\sqrt{-h})^2}\Big(- p_{\m\n}^2+\fr12  p^2\Big)=0 \la{const1}
\eea
\bea
\dot{h}_{\m\n} &=& -2n\fr1{\sqrt{-h}}(p_{\m\n}-\fr12 h_{\m\n}p)
   \la{gd}
\eea
where $G^{\m\n}\equiv R^{\m\n}-\fr12 h^{\m\n}R$ and $p\equiv h_{\m\n}p^{\m\n}$. 
The surface of a fixed metric still has the degrees of freedom associated with $n$ and, therefore, the entropy of the hypersurface with a fixed metric will naturally be associated with the dynamics of the field $n$.  
Also note that
\bea
p_{\m\n}=\sqrt{-h}\Big(-\fr1{2n}\pa_r h_{\m\n} +h_{\m\n} \fr1{2n}h^{\r\s}\pa_r h_{\r\s} \Big)
  \la{Kpi}
\eea
and
\bea
\fr1{\sqrt{-h}}(-p_{\m\n}^2+\fr12  p^2)=\sqrt{-h}\Big[-\fr1{4n^2}(\pa_r h_{\m\n})^2
     + \fr1{4n^2}(h^{\r\s}\pa_r h_{\r\s})^2\Big]
\eea
The Hamiltonian formulation of $r$-evolution \rf{hamiltonian} admits, as it should, the Schwarzschild metric as a
solution. One can check this explicitly as follows. Taking $n=n(r)$, the equation \rf{mtmdr} becomes
\bea
-\fr{1}{n\sqrt{-h}}\pa_r{p}^{\m \n} &=& \,G^{\m \n}+\fr12 \fr{1}{(\sqrt{-h})^2} (p_{\r\s}^2-\fr12 p^2)h^{\m \n}
  -\fr{2}{(\sqrt{-h})^2} (p_\r^\m p^{\n \r}-\fr12 p p^{\m \n})\nn\\ \la{mtmd2}
\eea
The trace part of the RHS of \rf{mtmd2} implies
\bea
h_{\m\n}\pa_r{p}^{\m \n}=0
\eea
This and the non-trace part of \rf{mtmd2} can easily be shown to lead to the solution of
\bea
n^2= \left(1-\fr{2GM}{r}\right)^{-1}
\eea
(The integration constant has been fixed in the usual manner.)
Once the solution for $h_{\m\n}$
is obtained, $p_{\m\n}$ can be determined by \rf{Kpi}.
The result implies that the Schwarzschild metric
\bea
ds^2=-\Big(1-\fr{2GM}{r}\Big)dt^2+\Big(1-\fr{2GM}{r}\Big)^{-1}dr^2+r^2(d\th^2+\sin^2\th d\varphi^2)
\eea
is indeed a solution of the $r$-evolution Hamiltonian.
For later use, we introduce the 3D metric solution, which we call ``the 3D Schwarzschild metric",
\bea
ds_{3D}^2 &=& -\Big(1-\fr{2GM}{r}\Big)dt^2+r^2(d\th^2+\sin^2\th d\varphi^2) \nn\\
          &=& -\Big(1-\fr{2GM}{r}\Big)dt^2+\fr{4r^2}{(1+z\zb)^2}dzd\zb
\la{hmunu}
\eea
{In the second equality}, the two-sphere part has been rewritten by introducing complex coordinates $z,\zb$.

\subsection{Metric rescaling and reduction \label{metric rescaling}}

One may consider the system given by \rf{L},  
and in principle attempt to analyze the fluctuations around 
the Schwarzschild solution. However, the quantum level analysis 
would be complicated because of the non-canonical manner in which the moduli field $n$ enters. 
In this section, we rescale the metric in order to make the system and its analysis more tractable.
Towards the end of this section, we will note the effective reduction to 2D; it is that 2D theory whose entropy is computed in section 3.

\vspace{.1in}
Let us consider the Lagrangian \rf{L}, and rescale the metric by
\bea
h_{\m\n} \rightarrow e^{\phh} h_{\m\n} \la{mr}
\eea
(More systematically, this step amounts to defining $h_{\m\n} = e^{\phh} \hat{h}_{\m\n}$ and omitting the hat in $\hat{h}$ for simplicity of notation.)
The Lagrangian now takes
\bea
\cL &=& n\sqrt{-h}\,e^{\fr32 \phh}\Big[ \fr{3}{2n}\pa_r \phh+\fr{1}{2n} h^{\m\n}\pa_r h_{\m\n} \Big]^2  
      -n{\sqrt{-h}} \,e^{-\fr12 \phh}\Big[\fr{1}{2n}(\pa_r \phh)e^{\phh}h_{\m\n} 
         +\fr{e^{\phh}}{2n}\pa_r h_{\m\n} \Big]^2  \nn\\
    &&+n e^{\fr12 \phh}\sqrt{-h}\Big[R^\3-2\N_\m\N^\m \phh-\fr12 \N_\m \phh\N^\m \phh \Big]                
\eea
Strictly speaking, there should be additional terms that arise from the virtual boundary effect. (See \cite{Park:2013vpa} and \cite{Park:2013bma} for more careful analysis of them.) We trace them more carefully in Appendix B where the details of the reduction are presented.
Once one chooses $\f$ (or $n$) such that
\bea
{e^{\phh}}=\fr1{n^2} 
\eea
one gets
\bea
\cL &=& \sqrt{-h}\,{e^{2 \phh}}\Big( \fr{3}{2}\pa_r \phh+\fr{1}{2} h^{\m\n}\pa_r h_{\m\n}
                \Big)^2  
      -{\sqrt{-h}} \,{e^{2 \phh}}\Big(\fr{1}{2}h_{\m\n}\pa_r \phh 
         +\fr{1}{2}\pa_r h_{\m\n} \Big)^2  \nn\\
    &&+\sqrt{-h}\Big[R^\3-2\N_\m\N^\m \phh-\fr12 \N_\m \phh\N^\m \phh \Big]
    \la{4d3dL}
\eea
Is is straightforward to obtain the scalar and metric field equations and they are presented in Appendix A. Let us carry out the reduction to 3D by taking the following ansatz
\bea
\hat{\f}(r,t,\th,\vf)&=& \f_0(r)+\f(t,\th,\vf) \la{rreduction}\;\;\mbox{with}\;\; e^{\f_0}\equiv \fr1{n_0^2} \nn\\
h_{\m\n} &=& h_{0\m\n}(r)
\la{fs}
\eea
where $n_0$ denotes 
\bea
n_0^2=\left(1-\fr{2GM}{r}\right)^{-1}
\eea
\begin{figure}
\centerline{
\begin{minipage}[b]{7cm}
             \epsfxsize=7cm
              \epsfbox{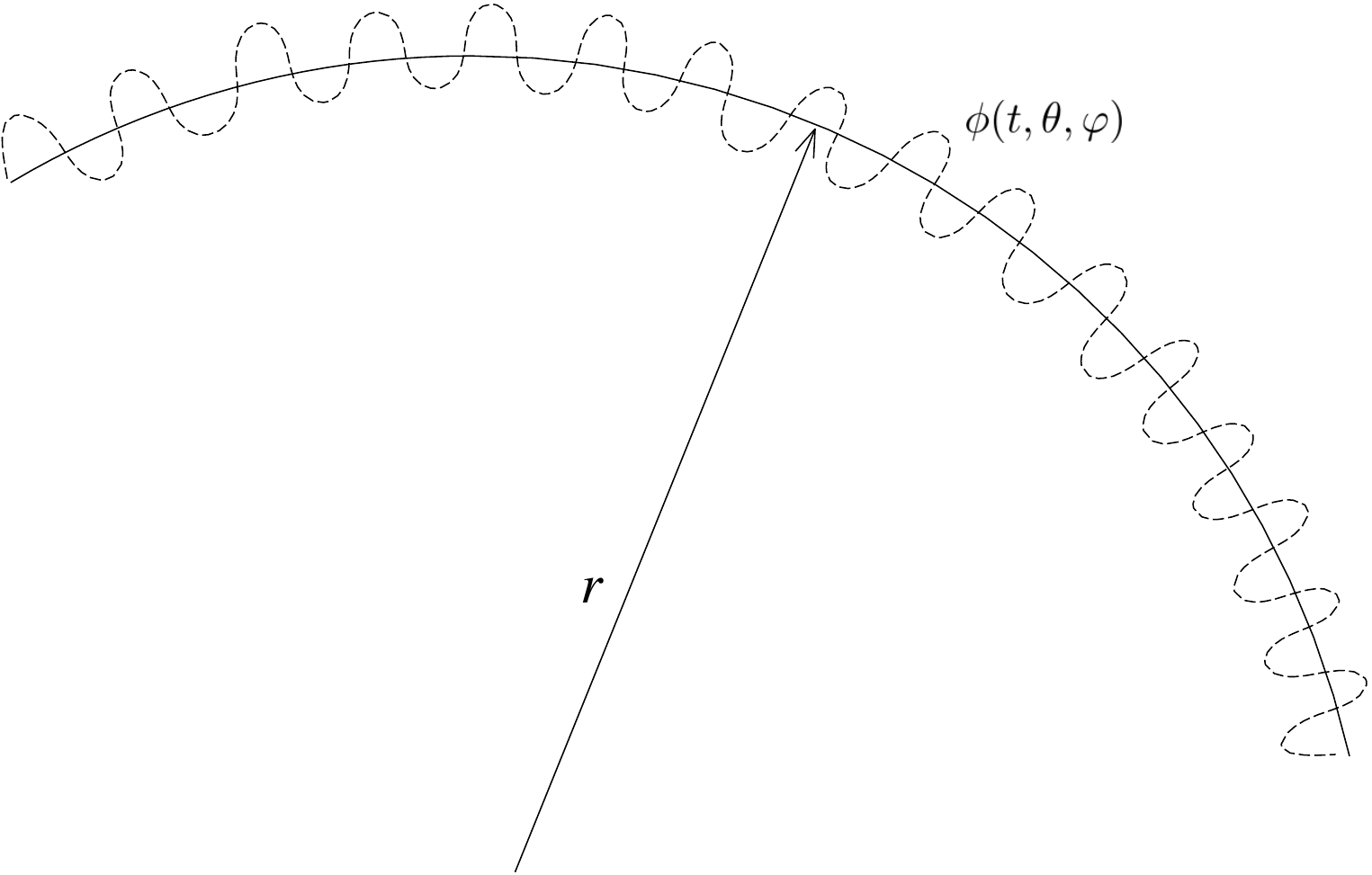}
      \end{minipage}
      }
\caption{reduction along $r$}
\label{fig}
\end{figure}
The functions that only have $r$-dependence may be treated as constants because the coordinate $r$ has become a parameter after the reduction.
(Remember that, by reduction, we have narrowed down to a sub-sector of the original 4D theory.)
Upon substituting \rf{fs} into the field equations, one gets 
\bea
&& \hspace{1in}\N^2 \f +\k(r) e^{2\f}=0 \nn\\
&&-\fr12 \pa_\m \f \pa_\n \f+\fr14 h_{0\m\n}(\pa \f)^2   +f_{\m\n} e^{2\f}+G_{0\m\n}=0   \la{3dVira}
\eea
where the explicit expressions for $\k,f_{\m\n}, G_{0\m\n}$ can be found in
Appendix A. As $r\ra R_S$, they become
\bea
\k\ra\fr{8}{R_S^2},\quad f_{\m\n}\ra0,\quad G_{0\m\n}\ra0
\eea
The second equation in \rf{3dVira} comes from the metric field equation, and can be viewed as a constraint for the $\f$ field in 3D.

In Appendix A, we construct a 3D action that produces the field equations \rf{3dVira} up to certain freedom associated with Liouville type theories
; we simply quote the result here:
\bea
\int d^3x\cL_{3D}= \int d^3x\sqrt{-h}\Big(
-\fr12 \pa_\m \f \pa^\m \f  
\Big)+\fr{\k}{2}\int d^3x \sqrt{\g}\; e^{2\f} 
\la{3dsaction3q} 
\eea
This action allows one to describe the physics of the surface of a fixed radius 
through the dynamics of $\f$.\footnote{A more proper step would 
be integrating out the fluctuating metric around the fixed metric in the path integral. 
Once the metric background is chosen to be the Schwarzschild background, {\em the theory is in that curved background}. This with the work of 
\cite{Hatefi:2012bp} confirms once again that the proposal put forward in \cite{Park:1999xz} was correct: the holographic dual theory should be placed in the curved background. (See, e.g., \cite{Park:2008fp} for further discussion.) \label{int}} It describes a branch of the moduli space of the hypersurface at an arbitrary
$r\geq R_S$ location.  
Let us consider the $\f$-fluctuations around the metric 
\bea
ds_{3D}^2&\equiv& h_{0\m\n}dx^\m dx^\n = -dt^2+{n_0^2} r^2(d\th^2+\sin^2\th d\varphi^2)
\eea
Once the hypersurface is identified with the event horizon, $r=R_S$, 
the time-dependent configurations give contributions that oscillate rapidly in the path-integral, and therefore produce a suppressing factor due to the presence of the factor $n_0$. In the Minkowski case, this is due to Lebesgue lemma. 
The same is true for the Euclidean case that results from $t \rightarrow it$.
 In section 3, we 
will consider the 2D action that results from this reduction.  
As stated in Appendix A, one can use the theory effectively reduced to 2D for the computation of the entropy:  
\bea
S_{2D} 
&=& \int d^2z \Big(-\pa_z \f\pa_{\zb} \f+{2}\r_0 e^{2 \f} \Big)
\la{liouvilletype0}
\eea
where
\bea
\r_0\equiv \fr{4}{(1+z\zb)^2}
\eea
Compared with the standard form of Liouville theory,\footnote{See, e.g., \cite{Ginsparg:1993is,Fateev:2000ik,Teschner:2001rv,Nakayama:2004vk} for a review. Liouville theory appeared in the usual Kaluza-Klein context, e.g., in \cite{Lu:1996hh}.} \rf{liouvilletype0} has several differences. First, the kinetic term and the potential term have opposite signs. (We will come back to this point in the next section.) The linear term in $\f$ is absent in
\rf{liouvilletype0}.\footnote{Liouville thoery without the linear term was discussed for example in \cite{Matone:1993tj}. We thank Yu Nakayama for pointing out this reference.}
As a matter of fact, there are several sources for linear terms. For example, a linear term appears once the factor $\r_0$ in front of the $e^{2\f}$ - which is another difference - is absorbed by shifting $\f\rightarrow \f+{\f}_0$ with an appropriate function ${\f}_0$:
\bea
S_{2D} 
&=& \int d^2z 
\Big(- \pa_z \f\pa_{\zb} \f-\pa_z \f\pa_{\zb}{\f}_0 - \pa_z {\f}_0\pa_{\zb} \f- \pa_z {\f}_0\pa_{\zb} \f_0+ e^{2 \f} \Big)\nn\\
&=& \int d^2 z 
\Big(- \pa_z \f\pa_{\zb} \f+2\f\pa_z \pa_{\zb}{\f}_0 - \pa_z {\f}_0\pa_{\zb} \f_0+ e^{2 \f} \Big)
\la{preliou}
\eea
where
\bea
e^{2\f_0}=\fr1{\r_0}  \la{phi0}
\eea
The linear term $\f\pa_z \pa_{\zb}{\f}_0$ will be addressed in the next section.
There is another contribution to the linear term. 
The field redefinition \rf{mr} should come with a determinant factor in the path integral over $h_{\m\n}$.
The determinant factor brings the $\f$-linear term with an infinite coefficient (see, e.g.,\cite{Tseytlin:1989si}). Therefore it is natural to expect that the linear term would arise through renormalization procedure even if it was absent in the beginning.

Although we will pursue the renormalization procedure elsewhere \cite{iyp} (since it deserves thorough 
work in its own right), we sketch what needs to be established in that procedure.
We find the issue of renormalization critical for several reasons. 
Firstly, the appearance of
the linear term with an infinite coefficient that comes from the measure should be a strong indication of the need for renormalization at a fairly early stage of the analysis.   
The second reason has something to do with the fact that we will only consider the kinetic term and the exponential term in \rf{preliou} in the next section.  
{The linear term coming from the measure must be addressed in the renormalization since it has an infinite coefficient.}
In terms of renormalization jargon, our setup amounts to starting with the bare action
\bea
S_{2D} 
&=& \int d^2x  \Big(-\fr{1}2 \pa_a \f_B\pa^a \f_B+Q_BR\f_B+\m_B e^{2b_B \f_B} \Big)
\la{bare}
\eea
and splitting it into the renormalized part and the counter terms:
\bea
S_{2D} 
&=& \int d^2x  \Big(-\fr{1}2 \pa_a \f_r\pa^a \f_r+\m_r e^{2b_r \f_r} \Big)+
\D S_{ctr}
\la{bare}
\eea
where the counter-term part contains the linear term. (A related discussion was given in \cite{Distler:1988jt}.)
Although we believe that such a renormalization program is very plausible, it must be explicitly established. We assume for now that this can be done which implies the presence of conformal symmetry.

\section{Horizon-area scaling entropy from $\f$-field}

In the previous section, we have carried out the reduction along the
radial (and time) direction(s) and obtained a Liouville type theory. 
It is actually (a version of) time-like Liouville theory due to the opposite signs
of the kinetic term and the potential term.

In this section we compute the partition function and entropy.
With the discussion of the previous subsection, we take 
\bea
S_{2D} 
&=& \int d^2z \Big(- \pa_z \f\pa_{\zb} \f+2\f\pa_z \pa_{\zb}{\f}_0 
     - \pa_z {\f}_0\pa_{\zb} \f_0+e^{2 \f} \Big)
\la{liouvilletype5}
\eea
for our starting point for the entropy 
computation. Here the subscript ``r" has been omitted and also in the following.
The terms containing ${\f}_0$ break the conformal symmetry, therefore will give the 
radius dependent effects once we transform the unit disc into a disc of larger radius.
Since the integration is over the unit disc, one would impose Dirichlet type boundary 
condition on $\f$. We postpone the systematic analysis for the future, 
and take a heuristic path below showing that the ${\f}_0$ containing fields should produce
only the large radius effects. 

Suppose the following rescaling
\bea
w= r_0 z\quad,\quad    \wb= r_0 \zb 
\eea
upon which
\bea
S_{2D} 
&=& \int d^2w\; \Big(- \pa_w \f\pa_{\wb} \f+2\f\pa_w \pa_{\wb}{\f}_0 
     - \pa_w {\f}_0\pa_{\wb} \f_0+e^{2 \f} \Big)
     \la{confbreakingact}
\eea
where now the arguments of $\f,\f_0$ are $\fr{w}{r_0}$: $\f=\f({w}/{r_0}),\f_0=\f_0({w}/{r_0})$. With this rescaling
\bea
e^{2{\f}_0}=\left[\fr{4}{(1+\fr{w\wb}{r_0^2})^2} \right]^{-1}\la{phitilde0}
\eea
which yields
\bea
{\f}_0=\fr12\ln \fr{(1+\fr{w\wb}{r_0^2})^2}{4} \la{phitilde1}
 =-\fr12 \ln4+ \fr{w\wb}{r_0^2}+\cdots
\eea
One may carry out perturbation in $\fr1{r_0^2}$.
Let us consider the leading order result in $r_0$ dropping the 2nd and 3rd terms in \rf{confbreakingact}. 
(The sub-leading terms may possibly affect
the terms of inverse powers of $R_S$ that are represented by the $(+\cdots)$ part in \rf{entropy}.) 
Let us now turn to the entropy computation of this 2D theory. 
After reduction of the radial and time directions, an overall volume factor will come from
\bea
\int d^4x \Rightarrow V_{(t,r)}\int d^2x\quad,\quad  V_{(t,r)}\equiv R_S^2\,\n
\eea
where the volume associated with $(t,r)$ directions has been defined in terms of the scale of the theory $R_S$ and a dimensionless number $\n$. 
This step is crucial for the derivation of the area scaling as we will see shortly. The 2D action \rf{liouvilletype5} becomes
\bea
S_{2D} 
=\int d^2z  \Big(-\fr{R_S^2}{G} \pa_z \f\pa_{\zb} \f+\fr{R_S^2}{G}\, e^{2 \f} 
   +2\fr{R_S^2}{G}\f\pa_z \pa_{\zb}\tilde{\f}_0 - \fr{R_S^2}{G}\pa_z \tilde{\f}_0\pa_{\zb} \tilde{\f}_0\Big)
\la{liouvilletype2}
\eea
By shifting $\f$ by an appropriate constant, the $\fr{R_S^2}{G}$ in front of the potential $e^{2\f}$
can be absorbed; one obtains
\bea
S_2 
&=& \int d^2z  \Big(-\fr{R_S^2}{G} \pa_z \f\pa_{\zb} \f+ e^{2 \f} 
 \Big)
\la{liouvilletype3}
\eea
Let us use this form of action in the following to compute the entropy of the $\f$-system.
Without this absorption of $\fr{R_S^2}{G}$ in front of $e^{2\f}$, the perturbation series would diverge: suppose one uses \rf{liouvilletype2} and insert several single-line vertices (i.e., the linear term), then the result will come with positive powers of $\fr{R_S^2}{G}$, and, therefore, the large-$R_S$ perturbation will break down.

Let us find a solution of the system; 
the field equation that follows from \rf{liouvilletype3} is
\bea
\pa_z\pa_{\zb} \f=- \fr{G}{ R_S^2}e^{2\f}\la{diffeq}
\eea
This admits the following solution \cite{Matone:1993tj}:
\bea
e^{2\f_c}=\fr{R_S^2}{ G}\fr{1}{(1+|z|^2)^2} \la{soliton}
\eea
Next we carry out perturbation around this solution and obtain the first two terms in the series for an illustration.
Since we are interested in partition function, our task is simpler than computing general 
correlators of the interacting Liouville theory.
Let us shift $\f\rightarrow \f_c+\f$ where $\f_c$ denotes the solution: 
\bea
S_{2D} 
&=& \int d^2z  \Big(-\fr{R_S^2}{G} \pa_z \f_c\pa_{\zb} \f_c-
\fr{R_S^2}{G} \pa_z \f\pa_{\zb} \f +e^{2 \f_c}(e^{2 \f}-2\f) \Big)\nn\\
&=& -\fr{R_S^2}{G}\int d^2z   \pa_z \f_c\pa_{\zb} \f_c+\fr{R_S^2}{G}\int d^2z
\left[ - \pa_z \f\pa_{\zb} \f+\fr{1}{(1+|z|^2)^2}(e^{2 \f}-2\f) \right]\nn\\
\la{liouvilletypeexp}
\eea
The classical piece gives
\bea
 -\fr{R_S^2}{G}\int {(2d^2 \s)}   \pa_z \f_c\pa_{\zb} \f_c
  &=&- \fr{R_S^2}{G}\left(  \pi \ln \fr{R_S^2}{G}
  +const \right)  \la{classic}
\eea
Although it may be possible to non-perturbatively evaluate the partition function by using the existing results in Liouville theory
literature, we ponder the perturbative method for now because it reveals interesting patterns.
The fluctuating part can be evaluated by the standard perturbation theory, which will produce a series
in the "coupling constant" $\fr{G}{R_S^2}$. Consider the fluctuation part,
\bea
S_{quam}=\int d^2z \left(- \fr{R_S^2}{G} \pa_z \f\pa_{\zb} \f+\fr{R_S^2}{G} \fr{1}{(1+|z|^2)^2}(e^{2 \f}-2\f)\right)
\la{fluc}
\eea
We may treat the exponential term of $(e^{2 \f}-2\f)= 2 \f^2+\cdots $ as vertices. The potential term has the ``wrong sign" - we will comment on the potential significance of the sign in the conclusion. 
As far as the kinetic term is concerned, one choice of regularization seems to stick out: putting the system 
on a sphere of unit radius followed by zeta function regularization. This seems to be a natural choice given that \rf{liouvilletypeexp} has originated from the 2D action on two-sphere. Whether this regularization will also be effective for dealing with the vertices remains to be seen.  
The leading piece for the fluctuating part comes from the determinant of the kinetic part.
The result depends on regularization method. (This seems to be a general feature of quantum entropy computations
 (see, e.g., \cite{Park:2012tba}).)
The 2D Laplacian is given by
\bea
\N^2& =& \left[\fr1{\sin\th}\fr{\pa}{\pa \th}\Big( \sin\th \fr{\pa}{\pa \th} \Big)
      +\fr1{\sin^2\th}\fr{\pa^2}{\pa \vf^2}\right]
\eea
and its eigenfunctions are the spherical harmonics:
\bea
\N^2 Y_{lm}=-l(l+1)Y_{lm}
\eea
One can use them to compute the determinant of the propagator. Focusing on the $R_S$-dependent part, one finds 
{after effective action (which is the analogue of the free energy) computation}
\bea
  \left[\det\left(\fr{G}{R_S^2}\N^2\right)\right]^{-1/2}\sim e^{-\fr12 c\tr \ln \fr{R_S^2}{G}}
\la{thecor2}
\eea
where only the $\fr{G}{R_S^2}$-dependent part has been kept
and the constant $c$ is defined by
\bea
c\equiv \sum_{l=0}^{\infty}(2l+1)=\fr13 \la{c}
\eea
where the factor $(2l+1)$ is due to the degeneracies of the eigenvalues. The zeta function regularization $\z(0)=-\fr12,\;\z(-1)=-\fr1{12}$ has been 
used in \rf{c}.
Let us combine \rf{classic} and \rf{thecor2}:
\bea
\exp\fr{R_S^2}{G}\left(-{ \pi}\ln \fr{R_S^2}{G}
 -\fr{c}2  \fr{G}{R_S^2} \ln \fr{R_S^2}{G} +\cdots  \right)
 \la{total}
\eea
where (...) denotes constant term and the terms of inverse powers of $\fr{R_S^2}{G}$ that come from the vertices.
Unlike in the finite temperature 
field theory setup, the entropy cannot be defined as the temperature derivative 
of the free energy since the temperature does not explicitly enter our zero-temperature 
setup. Noting that the coupling constant (or the square of the coupling constant to be precise) 
plays the role of temperature, a derivative of the free energy with respect to the coupling 
constant (or its square) was taken to be the entropy analogue in \cite{Hatefi:2012sy}\cite{Park:2012tba}. 
(See Appendix B for a more detailed discussion.)
In \rf{liouvilletype3}, $\fr{G}{R_S^2}$
plays the role of the coupling constant, and the entropy $s$ can be defined in the same spirit as \cite{Hatefi:2012sy}\cite{Park:2012tba}. The free energy analogue can be read off from \rf{total}:
\bea
F=-{ \pi}\ln \fr{R_S^2}{G}-\fr{c}{2} \fr{G}{R_S^2} \ln \fr{R_S^2}{G} +\cdots 
\eea
from which it follows 
\bea
s &=& \fr{\pa}{\pa (G/R_S^2)}\left[-{ \pi}\ln \fr{R_S^2}{G}
         -\fr{c}{2} \fr{G}{R_S^2} \ln \fr{R_S^2}{G} +\cdots \right]\nn\\
  &=& { \pi}\fr{R_S^2}{G}-\fr{c}2 \ln \fr{R_S^2}{G}+\cdots
   \la{entropy}
\eea
where (...) contains constant term and terms of inverse powers of $\fr{R_S^2}{G}$. 
The leading term corresponds to $\fr1{4G} A_{EH}$ where $A_{EH}$ denotes the area of the horizon. The coefficient of the second term above is different from those of \cite{Kaul:2000kf} and \cite{Sen:2012dw}. 
We comment on the possible reasons for this difference in the conclusion.

\section{Conclusion}

In this work, we have applied the ADM reduction scheme to the 4D Einstein-Hilbert action, and 
obtained the 3D system that describes the dynamics of a given hypersurface.
When the location of the hypersurface is set at the Schwarzschild radius, the 3D theory is 
effectively reduced to Liouville theory.  

There are several future directions.
This work establishes (up to renormalization) that the BH entropy comes from the 
2D Liouville theory that is associated with the hypersurface.
 Careful study of renormalization is needed to tie up some of the loose ends.
The potential term in \rf{fluc} is not bounded below; this should signify 
a certain instability. It will be interesting to see whether the instability could be associated with the evaporation of the black hole, and therefore the event horizon.
Further work is needed to understand the origin of the different coefficient of the logarithmic correction compared with the coefficient obtained, e.g., in \cite{Kaul:2000kf}. It should be possible to determine the coefficients of the terms of inverse powers of $\fr{R_S^2}{G}$ in \rf{entropy}, and for this a systematic analysis including the $\f_0$ containing terms will be necessary.

There are potentially several sources of corrections to the result \rf{entropy}.
Firstly, there may be non-perturbative corrections. Just as the multi-instanton sectors provide corrections to various quantities in a gauge theory, the same may happen to the present case. 
In terms of the bulk physics, the fluctuating hypersurface will generate gravitational waves. The gravitation waves will propagate along $r$ (and, depending on the symmetry, possibly other) direction(s), and taking them into account in the current “foliation setup” would require the collective dynamics of the leaves.  
It might be that somehow the degrees of freedom associate with the gravitational wave are entangled with the hypersurface and may provide additional corrections to the entropy.\footnote{We can say the following in regards to whether or not the contributions of the gravitational wave degrees of freedom to the entropy/path integral should be included: On one hand, their contribution may be factored out.
 The collective motions along $r$ might be analogous to the center of mass motions of an atom in atomic quantum mechanics. If indeed so, those degrees of freedom would be factored out and should not contribute to the intrinsic entropy associated with the hypersurface. This belief is based in part on the fact that the 2D theory has been obtained as a result of {\em consistent} reduction of the 4D theory, and therefore should provide a complete description of the physics for that moduli branch. 
 
On the other hand, it might be interesting to consider $\f_0=\f_0(t,r)$ in \rf{fs}. If one integrates out the quantum field $\f$, then one would be left with $\f_0$, and the resulting system would produce solitonic (,i.e., non-perturbative) contributions to the path integral. Therefore, there is a possibility that the radially propagating motion would be described by the soliton mode $\f_0(t,r)$. Determining whether this contribution should be added to the entropy under consideration seems to be a subtle issue. This issue, not entirely unrelated to the issue in the previous paragraph, would require future work.} It would be interesting to see whether these corrections would make the coefficient of the second term that appear in \rf{entropy} coincide with those of \cite{Kaul:2000kf}.

Another issue is temperature-related.
We used the zero-temperature field theory setup even though the black hole has Hawking temperature. (See, e.g., \cite{Frolov:1995xe} and \cite{Solodukhin:2011gn} for the finite temperature approach.)
As wellknown, the black hole temperature was an induced effect in the original paper by Hawking. 
Therefore it is natural to expect that temperature should enter our zero-temperature setup (e.g., the 2D scalar theory) through certain dynamical (or possibly kinematical) effects.   
 We have selected a hypersurface and did not consider 
``interactions" among the hypersurfaces. The role of the interactions of the neighboring hypersurfaces could be that they may induce finite temperature effects to the selected hypersurface.\footnote{One can see the justification of the zero-temperature setup by considering angular reduction. The 2D theory in the $(t,r)$ plane will have the following background metric:
\bea
ds^2=-\Big(1-\fr{2GM}{r}\Big)dt^2+\Big(1-\fr{2GM}{r}\Big)^{-1}dr^2
\eea
Since this geometry is that of a 2D  black hole, the temperature effect will arise in the usual manner even if one starts with the zero-temperature setup.
} We plan to report on some of these issues in the near future.

\vspace{1in}
\ni {\bf Acknowledgements}\\

\ni I thank A. Nurmagambetov and C. Rim for useful discussions. I also benefited from discussions
with J. Cho, N. Kim, Y. Myung and S. Nam.

\newpage

\renewcommand{\theequation}{A.\arabic{equation}}
 \setcounter{equation}{0}
  \section*{Appendix A: Reduction details}

In this appendix, we carry out the reduction from 4D to 3D in detail.
The resulting action \rf{3dsaction3} below is defined in $(t,\th,\vf)$ space. 
Consistency of ADM reduction generally requires addition of virtual boundary terms to the bulk action;
the relevance of the virtual boundary terms has been illustrated in \cite{Park:2013vpa}
for angular ADM reduction.\footnote{The possibility of treating black hole horizons as a boundary was proposed in \cite{Carlip:1999db} on more physical grounds. The virtual boundary terms under present consideration are motivated by consistency of the reduction, and they are required for angular reduction as well.} Many of the following steps are parallel to those in \cite{Park:2013v pa}.

We quote \rf{4d3dL}, the 1+3 split form of the 4D action,  here for convenience:
\bea
\cL &=& \sqrt{-h}\,{e^{2 \phh}}\Big( \fr{3}{2}\pa_r \phh+\fr{1}{2} h^{\m\n}\pa_r h_{\m\n}
                \Big)^2  
      -{\sqrt{-h}} \,{e^{2 \phh}}\Big(\fr{1}{2}h_{\m\n}\pa_r \phh 
         +\fr{1}{2}\pa_r h_{\m\n} \Big)^2  \nn\\
    &&+\sqrt{-h}\Big[R^\3-2\N_\m\N^\m \phh-\fr12 \N_\m \phh\N^\m \phh \Big]
    \la{4d3dLq}
\eea
where all the fields have dependence on the full coordinates $(t,r,\th,\vf)$.
Let us add the following virtual boundary term\footnote{Although the form of the virtual boundary term may appear ``fine-tuned", we stress the bottom line: The bulk action is defined up to the virtual boundary terms. The boundary terms, once added, bring consistency to the reduction without affecting the 4D bulk field equations.}
\bea
{ e\int d^4x\; \sqrt{\tilde{g}}\;\N_{\gt}^2 \;e^{\fr32\phh}} \la{dsphh}
\eea
where $e$ is a constant to be determined shortly and 
\bea
\gt_{\bar{\mu}\bar{\nu}}dx^{\bar{\mu}}dx^{\bar{\nu}}
&\equiv&  n^2 dr^2+\tilde{h}_{{\mu}{\nu}}dx^{{\mu}}dx^{{\nu}} \nn\\
\tilde{h}_{{\mu}{\nu}}dx^{{\mu}}dx^{{\nu}}&\equiv&
-dt^2+n_0^2r^2 \g_{ab}dx^adx^b \la{newmetrics}
\eea
$\N_{\gt}$ and $\g_{ab}(\th,\vf)$ denote the covariant derivative (with the connection constructed out of $\gt_{\bar{\mu}\bar{\nu}}$) and the 2D metric respectively.  
As a total derivative term, the term \rf{dsphh} does not contribute to the 4D bulk field equations. (In other words, we consider the bulk variations such that the boundary fields are inert.) However, they should affect the reduced 3D action. The boundary term \rf{dsphh}
will play an important role when we construct the reduced 3D action, a task that we will undertake below after discussing the 4D bulk field equations. The bulk scalar and metric field equations are
\bea
&& \N^2 \phh+2e^{2\phh}\Big[\Big( \fr{3}{2}\pa_r \phh+\fr{1}{2} h^{\r\s}\pa_r h_{\r\s}
                \Big)^2  
      -\Big(\fr{1}{2}h_{\r\s}\pa_r \phh 
         +\fr{1}{2}\pa_r h_{\r\s} \Big)^2\;\Big] \nn\\
&&+\pa_r  \Big[-3 e^{2\phh}\Big( \fr{3}{2}\pa_r \phh+\fr{1}{2} h^{\r\s}\pa_r h_{\r\s}
                \Big) 
      +e^{2\phh}h^{\r\s}\Big(\fr{1}{2}h_{\r\s}\pa_r \phh 
         +\fr{1}{2}\pa_r h_{\r\s} \Big)\;\Big]     =0 
         \la{scalareom} \nn\\
\eea
and
\bea
&& R_{\m\n}-\fr12 R h_{\m\n}-\fr12 \N_\m \phh \N_\n \phh+\fr14 h_{\m\n}(\pa \phh)^2
   \nn\\
&& -\fr12 e^{2\phh}h_{\m\n}\Big[\Big( \fr{3}{2}\pa_r \phh+\fr{1}{2} h^{\r\s}\pa_r h_{\r\s}\Big)^2    -\Big(\fr{1}{2}h_{\r\s}\pa_r \phh 
         +\fr{1}{2}\pa_r h_{\r\s} \Big)^2\;\Big]\nn\\
&&  +  e^{2\phh}   (\pa_r h_{\m\n})  \Big( \fr{3}{2}\pa_r \phh+\fr{1}{2} h^{\r\s}\pa_r h_{\r\s}\Big)+e^{2\phh}\pa_r \phh \Big(\fr{1}{2}h_{\m\n}\pa_r \phh 
         +\fr{1}{2}\pa_r h_{\m\n} \Big) \nn\\
&&+\fr1{\sqrt{-h}}h_{\a\m}h_{\b\n}\;\pa_r \Big[e^{2\phh}\sqrt{-h}h^{\a\b} \Big( \fr{3}{2}\pa_r \phh+\fr{1}{2} h^{\r\s}\pa_r h_{\r\s}\Big)  \Big]\nn\\
&& -\fr1{\sqrt{-h}}h_{\r\m}h_{\s\n}\;\pa_r \Big[e^{2\phh}\sqrt{-h}h^{\a\r}h^{\b\s}\Big(\fr{1}{2}h_{\a\b}\pa_r \phh 
         +\fr{1}{2}\pa_r h_{\a\b} \Big)  \Big]\nn\\
&&-  2e^{2\phh}    \Big(\fr{1}{2}h_{\m\b}\pa_r \phh +\fr{1}{2}\pa_r h_{\m\b} \Big)
          h^{\b\s}
         \Big(\fr{1}{2}h_{\n\s}\pa_r \phh +\fr{1}{2}\pa_r h_{\n\s} \Big)   =0 
         \la{metriceom}
\eea
The equations \rf{scalareom} and \rf{metriceom} are still four-dimensional; they are just written in the 1+3 split forms. Suppose carrying out the $r$-reduction with $r$ set at $r=R_S$. (A limiting process $r\ra R_S$ should be used when necessary.)

As noted in the main body, setting $r=R_S$ causes further reduction of the path-integral for entropy to 2D; let us explicitly carry out reduction to 2D. (After achieving this, we will construct the 3D action whose reduction produces the 2D action.)
The reduction to 2D can be carried out by taking
\bea
\hat{\f}(r,t,\th,\vf)&=& \f_0(r)+\f(\th,\vf) \la{rreduction}\;\;\mbox{with}\;\; e^{\f_0}\equiv \fr1{n_0^2} \nn\\
ds_{3D}^2&=&-dt^2+h_{ab}(t,\th,\vf)dx^adx^b
\la{fsq1}
\eea
where $a=1,2$ and
\bea
n_0^2=\left(1-\fr{2GM}{r}\right)^{-1}
\eea
The 2D metric can completely be gauge-fixed, which introduces the Virasoro-type 
constraint that arises from the original 4D metric field equation. Therefore, it is more convenient to execute all of these steps at one stroke by using the following ansatz   
\bea
&& \hspace{.3in}\hat{\f}(r,t,\th,\vf)= \f_0(r)+\f(\th,\vf) \la{rreduction} \nn\\
ds_{3D}^2&\equiv& h_{0\m\n}dx^\m dx^\n = -dt^2+{n_0^2} r^2(d\th^2+\sin^2\th d\varphi^2)
\la{fsq2}
\eea
instead of \rf{fsq1}.
Once the ansatz in \rf{fs} is substituted, the scalar field equation takes (it is still written in the 3D-index notation for convenience)
\bea
\N_\m \N^\m \f+\k(r) e^{2\f}=0  \la{eom1}
\eea
where
\bea
&& \k(r)= 2e^{2\f_0}\Big[\Big( \fr{3}{2}\pa_r \f_0+\fr{1}{2} h_0^{\r\s}\pa_r h_{0\r\s}\Big)^2    -\Big(\fr{1}{2}h_{0\r\s}\pa_r \f_0 
         +\fr{1}{2}\pa_r h_{0\r\s} \Big)^2\;\Big] \nn\\
&&+\pa_r  \Big[-3 e^{2\f_0}\Big( \fr{3}{2}\pa_r \f_0+\fr{1}{2} h_0^{\r\s}\pa_r h_{0\r\s}
                \Big) 
      +e^{2\f_0}h_0^{\r\s}\Big(\fr{1}{2}h_{0\r\s}\pa_r \f_0 
         +\fr{1}{2}\pa_r h_{0\r\s} \Big)\;\Big]   \nn\\       
\eea
and the metric field equation takes
\bea
&& -\fr12 \pa_\m \f \pa_\n \f+\fr14 h_{0\m\n}(\pa \f)^2   
+f_{\m\n}(r) e^{2\f}+G_{\m\n}(r)=0   \la{eom2}
\eea
where
\bea
G_{\m\n}(r)&=& R_{0\m\n}-\fr12 R_0 h_{0\m\n}  \nn\\
f_{\m\n}(r) &=&  -\fr12 e^{2\f_0}h_{0\m\n}\Big[\Big( \fr{3}{2}\pa_r \f_0+\fr{1}{2} h_0^{\r\s}\pa_r h_{0\r\s}\Big)^2    -\Big(\fr{1}{2}h_{0\r\s}\pa_r \f_0 
         +\fr{1}{2}\pa_r h_{0\r\s} \Big)^2\;\Big]  \nn\\ 
     &&  +  e^{2\f_0}   (\pa_r h_{0\m\n})  \Big( \fr{3}{2}\pa_r \f_0+\fr{1}{2} h_0^{\r\s}\pa_r h_{0\r\s}\Big)+e^{2\f_0}\pa_r \f_0 \Big(\fr{1}{2}h_{0\m\n}\pa_r \f_0 
         +\fr{1}{2}\pa_r h_{0\m\n} \Big) \nn\\
&&+\fr1{\sqrt{-h}}h_{0\a\m}h_{0\b\n}\;\pa_r \Big[e^{2\f_0}\sqrt{-h_0}h_0^{\a\b} \Big( \fr{3}{2}\pa_r \f_0+\fr{1}{2} h_0^{\r\s}\pa_r h_{0\r\s}\Big)  \Big]\nn\\
&& -\fr1{\sqrt{-h_0}}h_{0\r\m}h_{0\s\n}\;\pa_r \Big[e^{2\f_0}\sqrt{-h_0}h_0^{\a\r}h_0^{\b\s}\Big(\fr{1}{2}h_{0\a\b}\pa_r \f_0 
         +\fr{1}{2}\pa_r h_{0\a\b} \Big)  \Big]\nn\\
&&- 2 e^{2\f_0}    \Big(\fr{1}{2}h_{0\m\b}\pa_r \f_0 +\fr{1}{2}\pa_r h_{0\m\b} \Big)
          h_0^{\b\s}
         \Big(\fr{1}{2}h_{0\n\s}\pa_r \f_0 +\fr{1}{2}\pa_r h_{0\n\s} \Big)     
\eea
Substitution of 
\bea
\f_0 =\ln \left(1-\fr{2GM}{r}\right)
\quad,\quad
h_{0\m\n}=\left(
\begin{array}{ccc}
-1 & 0 & 0\\
&\\
0 & r^2 n_0^2 &0 \\
&\\
0 & 0 & r^2 n_0^2 \sin^2\th
\end{array}
\right) 
\eea
leads to 
\[
\k(r) = \fr{4(9G^2M^2-9GM r+2r^2)}{r^4} 
\]
\bea
f_{\m\n} &=&   \left(
\begin{array}{ccc}
-\fr{1}{r^2}\left(1-\fr{2GM}{r}\right) & 0 & 0\\
&\\
0 & 0 &0 \\
&\\
0 & 0 & 0
\end{array}
\right) \quad,\quad
G_{0\m\n}  =\left(
\begin{array}{ccc}
\fr{1}{r^2}\left(1-\fr{2GM}{r}\right) & 0 & 0\\
&\\
0 & 0 &0 \\
&\\
0 & 0 & 0
\end{array}
\right)  \nn\\
  \la{rdepcoeffsq}  
\eea
$f_{\m\n}$ and $G_{0\m\n}$ vanish upon setting $r=R_S$:
\bea
f_{\m\n}|_{r=R_S}=0\quad,\quad G_{0\m\n}|_{r=R_S}=0
\eea
and $\k(r)$ takes
\bea
\k|_{r=R_S}=\fr{8}{R_S^2};
\eea
The field equations now take\footnote{Strictly speaking, an additional boundary term is needed because the $\f$ field equation does not admit $\f=0$ as a solution. A $\f$-linear term with an appropriate coefficient can be added from the virtual boundary contribution. The same issue in the angular case was addressed in detail in \cite{Park:2013vpa} and \cite{Park:2013bma}. We will not repeat the analysis here because the extra term will not change the final result of the entropy, \rf{entropy}. }
\bea
&& \hspace{.4in}\N^2 \f +\k e^{2\f}=0 \nn\\
&&-\fr12 \pa_\m \f \pa_\n \f+\fr14 h_{0\m\n}(\pa \f)^2   =0   \la{3dViraq3}
\eea

In the remainder, we construct a 3D action that produces the field equations \rf{3dViraq3} up to the freedom associated with a Liouville type theory. (As a matter of fact this freedom was pointed out in the main body in the context of the 2D Liouville theory. Similar freedom should exist for a 3D Liouville type theory.) This will establish consistency of the $r$-reduction from 4D to 3D.

The second equation in \rf{3dViraq3} implies that the 3D action should take 
\bea
\int d^3x\cL_{3D}= \int d^3x\sqrt{-h}\Big(
-\fr12 \pa_\m \f \pa^\m \f  
\Big)
\la{3dsaction3part} 
\eea
But then this action is incompatible with the first equation in \rf{3dViraq3}.
The problem can be fixed by adding an appropriate virtual boundary term. Consider adding \rf{dsphh} to the original 4D action,
\bea
&&{e\int d^4x\; \sqrt{\tilde{g}}\;\N_{\gt}^2 e^{\fr32\phh}}
={e\int d^3x\; n\sqrt{\tilde{h}}\;g^{rr}\pa_r e^{\fr32\phh}} \nn\\
&=&{ e\int d^3x\; \sqrt{\tilde{h}}\;e^{\fr12 \phh}\pa_r e^{\fr32\phh}} 
={ \fr32 e \int d^3x\; \sqrt{\tilde{h}}\;(\pa_r \phh) e^{2\phh}} \nn\\
&=&{ \fr32 e {R_S}\int d^3x\; \sqrt{\g}\; e^{2\f}} 
\la{dsphh2}
\eea
In the third equality, the ansatz $\hat{\f}(r,t,\th,\vf)=\f_0(r)+\f(t,\th,\vf) $ has been used. Therefore the 3D action takes, after choosing $e=\fr{8}{3R_S^3}$,
\bea
\int d^3x\cL_{3D}= \int d^3x\sqrt{-h}\Big(
-\fr12 \pa_\m \f \pa^\m \f  
\Big)+\fr{4}{R_S^2}\int d^3x \sqrt{\g}\; e^{2\f} 
\la{3dsaction3} 
\eea
 Let us take the bulk metric $h_{\m\n}$ in \rf{3dsaction3part} (which also 
 appears in the first term in \rf{3dsaction3}) to be 
\bea
ds^2=-\r^2 dt^2+h_{ab}dx^adx^b
\eea
The bulk $\r$ and $h_{ab}$ field equations of \rf{3dsaction3} produces the metric field equation in
\rf{3dViraq3}. One can gauge-fix the bulk fields
\bea
\r=1\quad,\quad h_{0ab}={n_0^2} r^2(d\th^2+\sin^2\th d\varphi^2)
\eea
Finally the $\f$ field equation has an additional factor 
$\fr{\sqrt{\g}}{\sqrt{-h}}$ multiplied with $e^{2\f}$.
This factor can be absorbed by shifting $\f$. The shift then produces 
$\f$-linear terms in the action. They in turn can be absorbed by the freedom associated with renormalization of a Liouville type theory. (The main body has discussion on this in the context of the 2D Liouville theory.) The action \rf{3dsaction3} can further be reduced to 2D and, with a few rescalings, the resulting 2D action can be brought to the form in \rf{liouvilletype0}.

\renewcommand{\theequation}{B.\arabic{equation}}
 \setcounter{equation}{0}
  \section*{Appendix B: QFT Zero Temperature entropy}

Conventionally, entropy of a quantum field theory system is considered in the finite-temperature setup. 
It is clear that in general a quantum field theory system at zero temperature should have finite entropy in general. 
We used a certain notion of zero-temperature entropy in \cite{Hatefi:2012sy} \cite{Park:2012tba} and also in the present work. Although the notion is based on a close analogy with the finite temperature case, we discuss it here explicitly.\footnote{A few months after the first version of this work, the author became aware that the present definition of entropy is in the same spirit as that of
\cite{Callan:1994py}. }

We limit our discussion to cases in which an Euclidean path-integral that takes the form of
\bea
\int d\Phi e^{-\fr{S(\Phi)}{\l}}
\eea
where $\Phi$ is the field that represents the degrees of freedom of the system 
and $\l$ is the ``coupling constant". Let us set this to
\bea
\int d\Phi e^{-\fr{S(\Phi)}{\l}}=e^{-\fr{F(\Phi,\l)}{\l}} \la{classaction}
\eea
 $\l$ plays a role analogous to temperature 
in a finite temperature system, and $F$ is an analogue of free energy \cite{peskin}. Taking a derivative with
respect to $1/\l$, one gets
\bea
\fr{\int d\Phi\; S e^{-\fr{S(\Phi)}{\l}}}{\int d\Phi e^{-\fr{S(\Phi)}{\l}}}=F +\fr1{\l}\fr{\pa F}{\pa (1/\l)}
\eea
The left-hand side is the average of the Hamiltonian, and the equation can be rewritten as
\bea
<H>=F -{\l}\fr{\pa F}{\pa \l}
\eea
where $<H>$ denotes the energy of the system.
This implies that $\fr{\pa F}{\pa \l}$ should play a role analogous to 
the entropy of a finite-temperature system, $s$:
\bea
s\equiv \fr{\pa F}{\pa \l}
\eea

\newpage

\end{document}